\documentclass[apj]{emulateapj}

\usepackage{natbib}
\newcommand{\eg}{e.g., }

\newcommand{\Msun}{M_{\odot}}
\newcommand{\kms}{km~s$^{-1}$}

\newcommand{\Nifs}{$^{56}$Ni}

\newcommand{\Mej}{M_{\rm ej}}
\newcommand{\KE}{E_{\rm K}}
\def\gsim{\mathrel{\rlap{\lower 4pt \hbox{\hskip 1pt $\sim$}}\raise 1pt
\hbox {$>$}}}
\def\lsim{\mathrel{\rlap{\lower 4pt \hbox{\hskip 1pt $\sim$}}\raise 1pt
\hbox {$<$}}}

\def\ion#1#2{{\rm #1}~{\sc #2}}

\shorttitle{SN 2008D AT NEBULAR PHASES: SIDE-VIEWED BIPOLAR EXPLOSION}
\shortauthors{Tanaka et al.}
 
\begin{document}

\title{NEBULAR PHASE OBSERVATIONS OF THE TYPE I\lowercase{b} SUPERNOVA 2008D/X-RAY TRANSIENT 080109: SIDE-VIEWED BIPOLAR EXPLOSION \altaffilmark{1}}

\author{
Masaomi Tanaka\altaffilmark{2, 3}, 
Masayuki Yamanaka\altaffilmark{4},
Keiichi Maeda\altaffilmark{3},
Koji S. Kawabata\altaffilmark{4},
Takashi Hattori\altaffilmark{5},
Takeo Minezaki\altaffilmark{6}, 
Stefano Valenti\altaffilmark{7},
Massimo Della Valle\altaffilmark{8,9},
D. K. Sahu\altaffilmark{10},
G. C. Anupama\altaffilmark{10},
Nozomu Tominaga\altaffilmark{11,12},
Ken'ichi Nomoto\altaffilmark{3,2},
Paolo A. Mazzali\altaffilmark{13,14}, and
Elena Pian \altaffilmark{15}
}

\altaffiltext{1}{Based on data collected at Subaru Telescope, 
which is operated by the National Astronomical Observatory of Japan.}
\altaffiltext{2}{Department of Astronomy, Graduate School of Science, University of Tokyo, Bunkyo-ku, Tokyo, Japan; mtanaka@astron.s.u-tokyo.ac.jp}
\altaffiltext{3}{Institute for the Physics and Mathematics of the Universe, University of Tokyo, Kashiwa, Japan}
\altaffiltext{4}{Hiroshima Astrophysical Science Center, Hiroshima University, Higashi-Hiroshima, Hiroshima, Japan}
\altaffiltext{5}{Subaru Telescope, National Astronomical Observatory of Japan, Hilo, HI}
\altaffiltext{6}{Institute of Astronomy, School of Science, University of Tokyo, 2-21-1 Osawa, Mitaka, Tokyo 181-0015, Japan}
\altaffiltext{7}{Astrophysics Research Centre, School of Maths and Physics, Queen's University, Belfast, BT7 1NN, Northern Ireland, UK}
\altaffiltext{8}{Capodimonte Astronomical Observatory, Salita Moiariello 16, I-80131, INAF- Napoli, Italy}
\altaffiltext{9}{European Southern Observatory, Karl-Schwarzschild-Strasse 2, D-85748, Garching, Germany}
\altaffiltext{10}{Indian Institute of Astrophysics, II Block Koramangala, Bangalore 560034, India}
\altaffiltext{11}{Department of Physics, Konan University, Okamoto, Kobe, Japan}
\altaffiltext{12}{Optical and Infrared Astronomy Division, National Astronomical Observatory, Mitaka, Tokyo, Japan}
\altaffiltext{13}{Max-Planck Institut f\"ur Astrophysik, Karl-Schwarzschild-Strasse 2 D-85748 Garching bei M\"unchen, Germany}
\altaffiltext{14}{Istituto Naz. di Astrofisica-Oss. Astron., vicolo dell'Osservatorio, 5, 35122 Padova, Italy}
\altaffiltext{15}{Istituto Naz. di Astrofisica-Oss. Astron., Via Tiepolo, 11, 34131 Trieste, Italy}

\begin{abstract}
We present optical spectroscopic and photometric observations 
of supernova (SN) 2008D, 
associated with the luminous X-ray transient 080109,
at $>300$ days after the explosion (nebular phases).
We also give flux measurements of emission lines from 
the \ion{H}{ii} region at the site of the SN, 
and estimates of the local metallicity.
The brightness of the SN at nebular phases is consistent 
with the prediction of the explosion models with an ejected
\Nifs\ mass of 0.07 $\Msun$, 
which explains the light curve at early phases.
The [\ion{O}{i}] line in the nebular spectrum shows a double-peaked 
profile while the [\ion{Ca}{ii}] line does not.
The double-peaked [\ion{O}{i}] profile
strongly indicates that SN 2008D is an aspherical explosion.
The profile can be explained by a torus-like distribution 
of oxygen viewed from near the plane of the torus.
We suggest that SN 2008D is a side-viewed, bipolar explosion 
with a viewing angle of $> 50^{\circ}$ from the polar direction.
\end{abstract}


\section{Introduction}
\label{sec:intro}

On 2008 January 9, a luminous X-ray transient was 
serendipitously discovered in NGC 2770 
during the follow-up observation of SN 2007uy 
in the same galaxy with the {\it Swift} satellite \citep{berger08}. 
An optical counterpart was also discovered at the position of 
the transient \citep{deng08,valenti0808D},
and it was named supernova (SN) 2008D \citep{liw08}.

The total energy emitted at X-ray wavelengths 
is $\sim 2 \times 10^{46}$ ergs, 
smaller than that of long gamma-ray bursts (GRBs) 
by a factor of $>1000$ \citep{soderberg08}.
The origin of the X-ray emission is being debated.
\citet{soderberg08}, \citet{chevalier08}, and 
\citet{katz09} interpreted 
the X-ray transient as a SN shock breakout.
On the other hand, \citet{xu08}, \citet{lixl08} and 
\citet{mazzali08} suggested that the transient 
is the least energetic end of GRBs or X-ray flashes.

SN 2008D is classified as Type Ib because of the 
presence of He lines \citep{soderberg08,mazzali08,malesani09,modjaz0808D}
while SNe associated with GRBs are all Type Ic (without He lines).
The progenitor star of SN 2008D 
has a He layer prior to the explosion, 
and the He core mass is estimated to be 6-8 $\Msun$ \citep{tanaka0908D}.

In this paper, we present optical spectroscopic and 
photometric observations of SN 2008D at $>$ 300 days after the 
explosion (nebular phases)
with the Subaru telescope equipped with FOCAS \citep{kashikawa02}
and the Very Large Telescope (VLT) equipped with FORS1 \citep{appenzeller98}.
We find that the spectrum of SN 2008D shows 
a double-peaked [\ion{O}{i}] emission profile,
suggesting that SN 2008D is a bipolar explosion 
viewed from near the equatorial direction.

\section{Observations and Data Reduction}
\label{sec:obs}

\begin{figure}
\begin{center}
\includegraphics[scale=0.65]{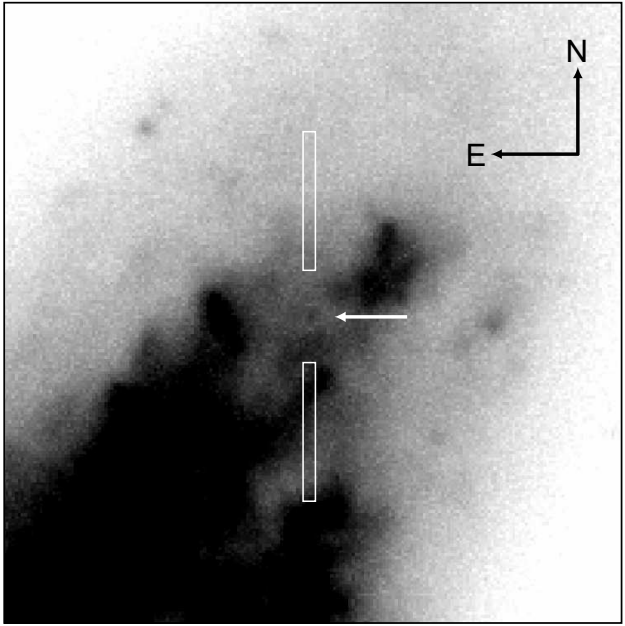}
\caption{
$40\farcs 0 \times 40\farcs 0$ section of the $R$-band image centered
on SN 2008D taken with Subaru telescope at $t_{\rm exp}=362.9$ days.
The position of the SN is shown with the arrow.
Two rectangles show the regions where the galaxy spectrum is 
extracted.
\label{fig:image}}
\end{center}
\end{figure}

\subsection{Spectroscopy}
\label{sec:spectroscopy}

A spectroscopic observation of SN 2008D was performed
on 2009 January 6 ($t_{\rm exp}=363$ days) with the Subaru telescope.
Hereafter $t_{\rm exp}$ denotes the 
epoch in observer's frame measured from the explosion date, 
MJD=54474.56 \citep{malesani09,modjaz0808D}.

Blue (4800-8000 \AA) and red (6000-9000 \AA) spectra were
taken separately.
We used a slit of $0\farcs 8$ width and two 300 lines mm$^{-1}$ grisms.
The slit was placed with the position angle of $0^{\circ}$ (north-south).
Typical seeing during the observation was FWHM $\simeq 1\farcs 2$ measured
with star profiles in $R$ band.
No filter was used for the blue spectrum,
while the O58 filter was used for the red spectrum
to eliminate second order light.
The exposure time is 9000 and 3300 sec for the blue
and red spectra, respectively.
The wavelength resolution is $\lambda/\Delta \lambda \sim 650$.

The observed data were bias-subtracted and flat-corrected,
and then two-dimensional spectra were extracted.
Wavelength calibration was performed with a Th-Ar lamp.

Since the SN is located inside a brightness trough which
surrounds the SN position with a diameter of $\sim 4$ arcsec
(see Figure \ref{fig:image}),
simultaneous subtraction of the background galaxy and sky lines
using the neighboring region leads to negative SN flux.
Thus, we first extracted a one-dimensional galaxy spectrum
using the region at 3-12 arcsec north and south from the SN position
(marked with boxes in Figure \ref{fig:image}).
Then, the galaxy spectrum was subtracted 
using the spatial profile at 6920-6990 \AA, 
in which the SN does not show any emission line.

To maximize S/N of the SN spectra especially at emission lines, 
we extracted the galaxy spectrum using a wide region.
Since the global color of the galaxy spectrum
depends on the position somewhat,
this leads to a slight over-subtraction in the extracted SN spectrum
in the blue (see Figure \ref{fig:spec}).
However, 
the profile of the SN emission lines are not affected by the 
choice of integrated region for the galaxy spectrum.

The background-subtracted spectra were summed 
into one-dimensional spectra, and then,
the flux was calibrated using the 
spectrophotometric standard star Feige 34 \citep{oke90}.
The blue and red spectra were combined,
weighted by the exposure time.
Finally, the spectrum is scaled with the $R$-band magnitude
(Section \ref{sec:photometry}).

\begin{deluxetable}{lr} 
\tablewidth{0pt}
\tablecaption{Observed flux of narrow emission lines}
\tablehead{
Emission line &
flux\tablenotemark{a} 
}
\startdata
H$\beta$ $\lambda$4861              & 9.42$\pm$ 0.68 \\
$[$\ion{O}{iii}$]$ $\lambda$ 5007   & 1.97 $\pm$  0.61\\
$[$\ion{N}{ii}$]$ $\lambda$ 6548    & 2.32 $\pm$ 0.66 \\ 
H$\alpha$ $\lambda$ 6563            & 33.1 $\pm$ 0.9 \\
$[$\ion{N}{ii}$]$ $\lambda$ 6584    & 7.00$\pm$ 1.27 \\ 
$[$\ion{S}{ii}$]$ $\lambda$ 6717    & 4.69 $\pm$ 0.52\\
$[$\ion{S}{ii}$]$ $\lambda$ 6731    & 3.18 $\pm$ 0.55\\
\enddata
\tablenotetext{a}{in units of $10^{-17}\ {\rm erg\ s^{-1}\ cm^{-2}}$.
Extinction is {\it not} corrected for.}
\label{tab:HIIregion}
\end{deluxetable}

\begin{deluxetable*}{lllrrrrrl} 
\tablewidth{0pt}
\tablecaption{Log of Photometric Observations}
\tablehead{
Date & 
MJD & 
Epoch \tablenotemark{a} & 
$m_B$ & $m_V$ & $m_R$ \tablenotemark{b} & $m_I$ & $M_{\rm bol}$ &
Telescope
}
\startdata
2008 November 22 & 54792.5 & 317.9 &     --                     & $>$22.4\tablenotemark{c}  & $23.0^{+0.5}_{-0.4}$  (22.7 $\pm$ 0.3)   &  22.4 $\pm$ 0.5  &  $-10.0 \pm 0.8$ & Subaru \\
2009 January 6  & 54837.5 & 362.9 &  $>$22.8\tablenotemark{c} & $>$22.2\tablenotemark{c}    & $23.8^{+0.9}_{-0.6}$  (23.2 $\pm$ 0.4)   & $>$22.1\tablenotemark{c}  & $-9.8 \pm 1.2 $ & Subaru \\ 
2009 February 19 & 54881.2 & 406.6 &    --                      &  $>$23.6\tablenotemark{c} & (23.9 $\pm$ 0.4)                      & $>$22.9\tablenotemark{c} &  --      & VLT \\
\enddata
\tablenotetext{a}{Days after the explosion (MJD=54474.56) in the observer's frame.}
\tablenotetext{b}{Value in parenthesis is total brightness without correction of the \ion{H}{ii} region.}
\tablenotetext{c}{5$\sigma$ upper limit.}
\label{tab:phot}
\end{deluxetable*}

\subsection{Narrow emission lines from the \ion{H}{ii} region}
\label{sec:HII}

There are very strong narrow emission lines from the 
\ion{H}{ii} region at the position of the SN.
Because of the faintness of the SN at nebular phases, 
the flux of the narrow emission lines can be easily measured.
Table \ref{tab:HIIregion} shows measured flux of the emission lines.
Extinction is {\it not} corrected for.

From the line ratios, we can estimate gas metallicity at
the site of SN 2008D.
\citet{thoene08} presented metallicity measurements 
using the data taken soon after the explosion.
We use N2 and O3N2 indexes to estimate the metallicity
\citep{pettini04}:
N2 $\equiv$ log([\ion{N}{ii}] $\lambda$6584/ H$\alpha$) 
and O3N2 $\equiv$ 
log( ([\ion{O}{iii}] $\lambda$5007/ H$\beta$)/
([\ion{N}{ii}] $\lambda$6584/ H$\alpha$) ).
Note that these indexes are almost independent on the extinction
because of the vicinity of the lines ([\ion{N}{ii}] and H$\alpha$, 
[\ion{O}{iii}] and H$\beta$).
\citet{pettini04} give the following relations calibrated with
the metallicity estimated by $T_e$ method:
12 + log(O/H) = 9.37 + 2.03$\times$N2 + 1.26$\times$N2$^2$ + 
0.32$\times$N2$^3$ and 8.73 $-$ 0.32$\times$O3N2, for N2 and O3N2 indexes, 
respectively.

Using these relations, 
metallicity is estimated to be 12+log(O/H) = $8.5 \pm 0.2$ (N2) 
and $8.7 \pm 0.2$ (O3N2).
The error above includes the 1$\sigma$ dispersion in the relation
(0.18 and 0.14 for N2 and O3N2, respectively).
These are consistent with the solar metallicity 
(12+log(O/H)=8.66, \citealt{asplund04}) within the error,
but larger than 
those by \citet{thoene08} by $\sim$0.2dex (with the same indexes).
The host galaxies of the GRB-associated SNe 
(GRB 030329/SN 2003dh, GRB 031203/SN 2003lw, XRF 020903, XRF 060218/SN 2006aj) 
have 12 + log(O/H) $\lsim 8.1$ in O3N2-$T_e$ scale \citep{modjaz08host}.
Thus, the metallicity at the site of SN 2008D is higher than
that in those environments, as pointed by \citet{thoene08}.

\begin{figure}
\begin{center}
\includegraphics[scale=0.95]{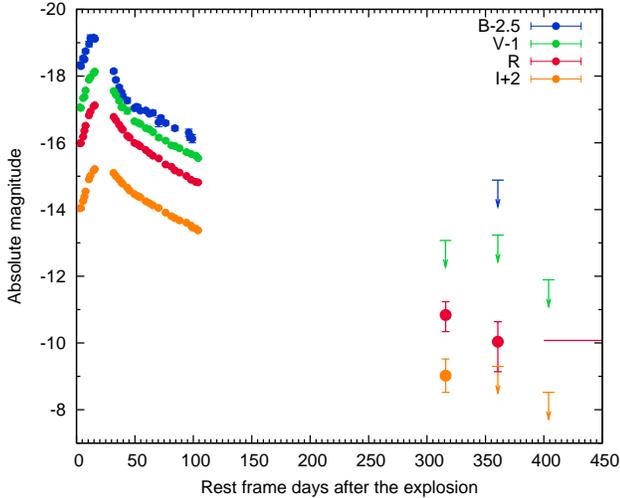}
\caption{
Reddening corrected, absolute $BVRI$ light curves of SN 2008D.
The red horizontal line shows estimated 
brightness of the \ion{H}{ii} region in $R$ band.
\label{fig:LC}} 
\end{center}
\end{figure}

\subsection{Photometry}
\label{sec:photometry}

Photometric observations in $BVRI$ were performed on 
2008 November 22 ($t_{\rm exp}=317.9$ days), 2009 January 6
($t_{\rm exp}=362.9$ days, Figure \ref{fig:image}),
and February 19 ($t_{\rm exp}=406.6$ days).
We detected a point source at the SN position 
in the $R$ band at three epochs and the $I$ band at one epoch.
A log of the observations is shown in Table \ref{tab:phot}.
We performed PSF photometry for the $R$- and $I$-band images,
where a source is detected.
For the images in which the SN was not detected,
limiting magnitudes were derived using 
the local sky noise around the SN position, 
and further checked using artificial sources.

Since the locally strong emission of the \ion{H}{ii} region 
is present at the SN position,
our $R$-band photometry overestimates the SN brightness.
We have corrected for this effect 
by estimating the contribution of the narrow lines 
using the observed spectrum at $t_{\rm exp}=363$ days 
($m_{R, \rm{gal}} \approx 24.1$ mag, 
the red horizontal line in Figure \ref{fig:LC}).
The error in the SN magnitudes at $t_{\rm exp}=317.9$ and $362.9$ days 
does not include the uncertainty caused by the contamination 
from galaxy light.
For the data at $t_{\rm exp}=406.6$ days, 
we refrain from measuring the SN component
because of the large uncertainty in the total flux 
as well as in the galaxy contribution.

\begin{figure}
\begin{center}
\includegraphics[scale=0.95]{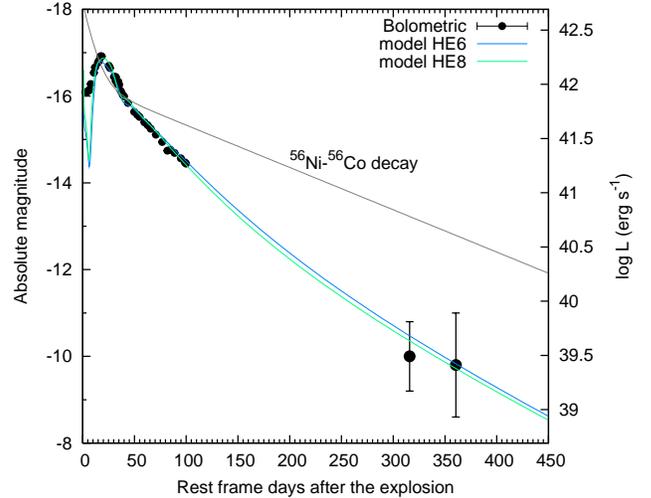}
\caption{
Pseudo-bolometric light curve (black) of SN 2008D.
The bolometric magnitudes are derived from $UBVRIJHK$ magnitudes at
early phases ($t_{\rm exp}<100$ days, T. Minezaki, in preparation) 
while they are estimated using the bolometric correction
for SN 2002ap at late phases \citep{tomita06}.
The solid lines in blue and green show the results of 
radiative transfer calculation by \citet{tanaka0908D}.
The gray line shows the total energy release from 0.07 $\Msun$ of \Nifs.
\label{fig:LCbol}} 
\end{center}
\end{figure}

\section{Light Curves Until Nebular Phases}
\label{sec:LC}

Figure \ref{fig:LC} shows the absolute magnitudes of SN 2008D.
For the reddening, $E(B-V)=0.65$ mag is assumed 
\citep{mazzali08,modjaz0808D},
and the reddening law of \citet{cardelli89} is adopted.
For the distance to the SN, $\mu=32.46$ is assumed \citep{modjaz0808D}.
The data at early phases ($t_{\rm exp} \lsim 100$ days)
are taken from T. Minezaki et al. (in preparation).

We estimate the bolometric luminosity of SN 2008D 
at nebular phases.
Since multicolor observations are not available at 
nebular phases, we use the bolometric correction of 
the well-observed SN 2002ap \citep{yoshii03,tomita06}.
We define the bolometric correction as
BC $\equiv M_{\rm bol} - M_{R, {\rm cor}}$, 
where $M_{\rm bol}$ is the bolometric magnitude
constructed from $UBVRIJHK$ magnitudes, 
and $M_{R, {\rm cor}}$ is the reddening corrected, 
absolute $R$-band magnitude (as plotted in Figure \ref{fig:LC}).
The BC of SN 2008D is found to be consistent with that of SN 2002ap
within 0.3 mag at early phases ($BC \sim 0.2-0.7$ mag).
The BC of SN 2002ap at nebular phases 
are applied to SN 2008D by interpolating the epoch.
The estimated BC for SN 2008D is 
$0.95 \pm 0.08$ and $0.75 \pm 0.2$ mag
at $t_{\rm exp}=317.9$ and 362.9 days, respectively.
For the bolometric magnitudes, 
a systematic uncertainty of 0.3 mag has been added.

The bolometric light curve is shown
in Figure \ref{fig:LCbol}.
The models by \citet[][blue and green lines]{tanaka0908D},
with an ejected \Nifs\ mass of 0.07 $\Msun$, are shown for comparison.
These models reproduce the bolometric 
light curve at early phases.
The bolometric magnitudes at nebular phases 
are consistent with the prediction of these models.
Ejecta mass and kinetic energy are different in two models
[($\Mej$/$\Msun$, $\KE$/$10^{51}$erg) = (4.4, 3.7) and (6.2, 8.4)
for model HE6 and HE8, respectively].
However, since the difference in the predicted luminosity
is very small, our observations at nebular phases 
do not discriminate between these two models.

\begin{figure}
\begin{center}
\includegraphics[scale=1.2]{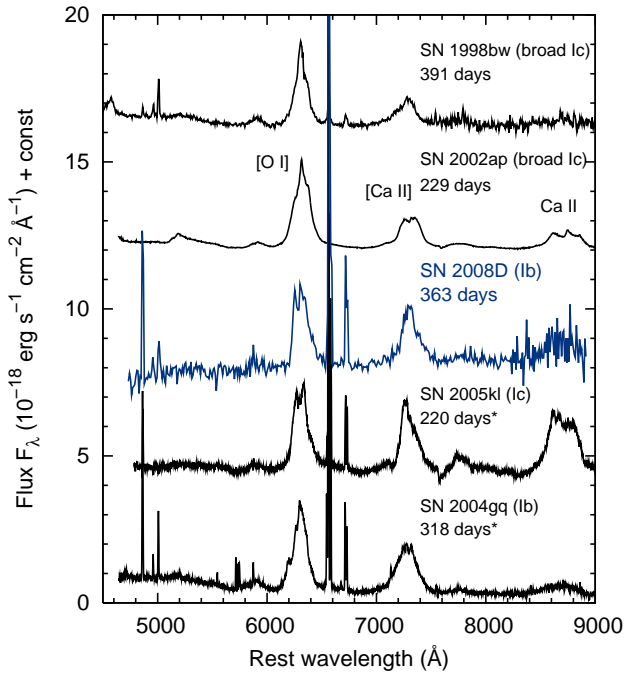}
\caption{
Nebular spectrum of SN 2008D (blue line) compared with 
that of SNe 1998bw \citep[broad Ic,][]{patat01}, 
2002ap \citep[broad Ic,][]{mazzali07}, 
2005kl \citep[Ic,][]{maeda08} and 2004gq \citep[Ib,][]{maeda08}.
The epochs are given as days since the explosion
(days since the discovery are given for the two epochs marked with $*$
as light curves are not available).
The spectra are normalized to the 
peak flux of the [\ion{O}{i}]$\lambda \lambda 6300,6364$ line
of SN 2008D.
The spectra are shifted by 16, 12, 8, 4, and 0 
from top to bottom for clarity.
\label{fig:spec}}
\end{center}
\end{figure}

\section{Nebular Spectrum}
\label{sec:nebular}

Figure \ref{fig:spec} shows the nebular spectrum of SN 2008D (blue line)
taken at $t_{\rm exp}=363$ days after the explosion.
The spectrum is binned into 10 \AA, 
similar to the wavelength resolution at $6300$ \AA.
It is compared with nebular spectrum of 
SNe 1998bw \citep[broad Ic,][]{patat01},
2002ap \citep[broad Ic,][]{mazzali07}, 2005kl \citep[Ic,][]{maeda08}, 
and 2004gq \citep[Ib,][]{maeda08}.
The spectrum of SN 2008D clearly shows emission lines of 
[\ion{O}{i}]$\lambda\lambda\ 6300, 6364$ and 
[\ion{Ca}{ii}]$\lambda\lambda\ 7291, 7323$.
These lines are commonly seen in the nebular spectra of other Type Ib/c SNe.
The \ion{Ca}{ii} IR triplet is marginally detected around 8600 \AA.
Narrow lines at 4800-5000 \AA\ and 6500-6900 \AA\ originate in
the \ion{H}{ii} region of the host galaxy.

\begin{figure}
\begin{center}
\includegraphics[scale=1.4]{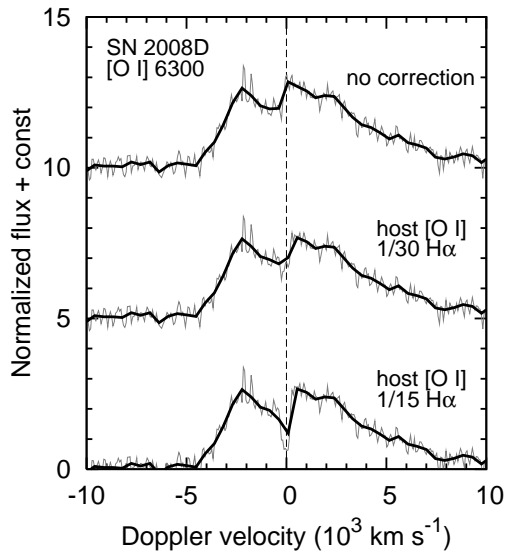}
\caption{
The [\ion{O}{i}] line profile of SN 2008D and 
possible contamination by the narrow [\ion{O}{i}] line from the \ion{H}{ii}
region.
{\it Top}: line profile extracted as in Section \ref{sec:spectroscopy} 
without correction of a possible narrow [\ion{O}{i}] line from the
\ion{H}{ii} region.
{\it Middle} and {\it bottom}: line profiles corrected for a possible 
narrow [\ion{O}{i}] line. 
The strength of the narrow [\ion{O}{i}] line is assumed to be 
1/30 and 1/15 of the H$\alpha$ line, respectively.
The gray lines show unbinned spectra while the black lines show 
the spectra binned into 10 \AA.
\label{fig:velOI}}
\end{center}
\end{figure}

\subsection{Line profile}

The [\ion{O}{i}] line of SN 2008D shows a double-peaked profile.
However, given the strong narrow emission lines from the \ion{H}{ii} region,
the broad [\ion{O}{i}] line of the SN could possibly be contaminated 
by the narrow [\ion{O}{i}] line from the \ion{H}{ii} region.
Since the strength of the narrow [\ion{O}{i}] line 
is $\sim 0 - 15$ of H$\alpha$ around the SN position, 
we subtract the scaled, gaussian-fitted H$\alpha$ profile 
at the position of [\ion{O}{i}].
Figure \ref{fig:velOI} shows the original profile 
(no correction, {\it top}) compared with
the profiles corrected with 1/30 and 1/15 of the H$\alpha$ line 
({\it middle} and {\it bottom}, respectively).
Although the peak at $v \sim 0$ \kms\ in the original spectrum 
may be due to the contamination of the narrow emission line,
the broad, double-peak profile is not affected by the contamination.

The [\ion{O}{i}] line shows a double-peaked profile,
while such clear two peaks are not seen in the [\ion{Ca}{ii}] line.
Although these properties were also seen in the spectrum 
taken at $t_{\rm exp}=109$ days \citep{modjaz0808D}, 
transparency of the ejecta was not sure 
at such a transition epoch from photospheric 
to nebular phases.
In fact, evolution of the profile from $t_{\rm exp}=109$ days 
is not significant.
Possible change can be seen in the redder peak of the [\ion{O}{i}] line.
At $t_{\rm exp}=109$ days it is clearly located at $v \sim 0$ \kms,
but the position can be redder at $t_{\rm exp}=363$ days, 
depending on the contamination of narrow [\ion{O}{i}] line from the 
\ion{H}{ii} region.

The [\ion{O}{i}] line profile is compared with other SNe 
in the left panel of Figure \ref{fig:line}.
Double-peaked profiles are also seen in other Type Ib/c 
SNe \citep{sollerman98,mazzali05,valenti0803jd,maeda08,modjaz08}.
The fraction of SNe showing a double-peaked [\ion{O}{i}] line 
is roughly $40 \pm 10 \%$ \citep{maeda08}.

Since our observation is late enough for the ejecta to be
optically thin, the double-peaked profile of the [\ion{O}{i}] line
is unlikely to be caused by optical depth effects
\citep{taubenberger09}.
It is also unlikely to be caused by a combination of the two 
[\ion{O}{i}] lines ($\lambda \lambda$ 6300 and 6364)
since the strength of the two peaks is comparable 
(the strength ratio is 3:1 in optically thin limit, \citealt{leibundgut91}).

In addition, if the double-peaked [\ion{O}{i}] profile were caused
by asphericity in the distribution of the heating source (\Nifs), 
the [\ion{Ca}{ii}] line would also show a double-peaked profile.
Thus, the profile of [\ion{O}{i}] line reflects the distribution
of excited \ion{O}{i}.
The [\ion{O}{i}] and [\ion{Ca}{ii}] lines may arise from different 
sites \citep{fransson89}.

\subsection{Comparison with the model}

\begin{figure}
\begin{center}
\begin{tabular}{cc}
\includegraphics[scale=1.1]{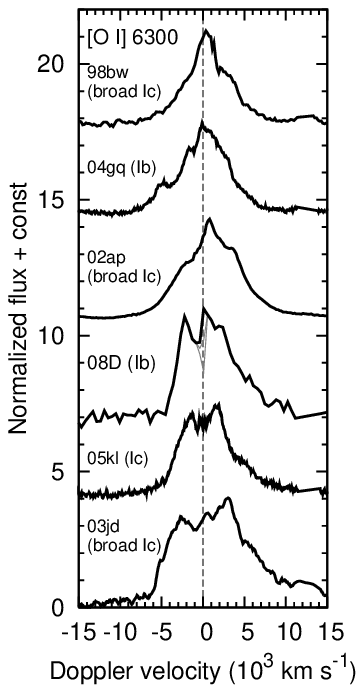} &
\includegraphics[scale=1.1]{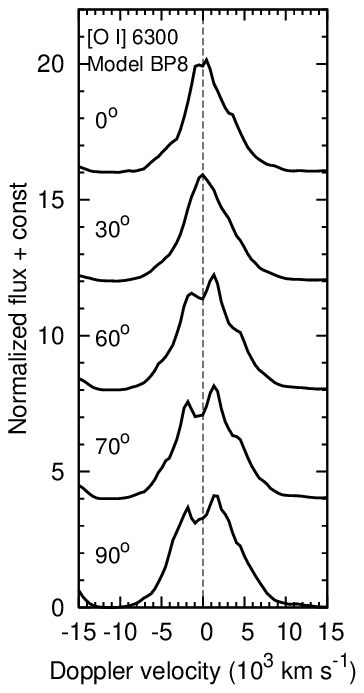}
\end{tabular}
\caption{
{\it Left:} 
Profiles of the [\ion{O}{i}] $\lambda \lambda 6300, 6364$ lines
plotted versus Doppler velocity measured from 6300 \AA.
The spectra are the same as Figure \ref{fig:spec},
but SN 2003jd \citep[broad Ic, 330 days][]{mazzali05,valenti0803jd} 
is added.
SNe 1998bw, 2004gq, and 2002ap show single-peaked
profile while SNe 2008D and 2005kl, and 2003jd
show double-peaked profile.
The gray line for SN 2008D shows the spectrum corrected
for the possible narrow [\ion{O}{i}] contamination 
(Figure \ref{fig:velOI}).
The spectra are shifted by 17.5, 14.0, 10.5, 7.0, 3.5, 
and 0.0 from top to bottom.
Narrow lines from the host galaxies are removed.
{\it Right:}
Profiles calculated with model BP8 by \citet{maeda06nebular,maeda08}.
The number represents the viewing angle 
measured from the polar direction.
The spectra are shifted by 16, 12, 8, 4, 
and 0 from top to bottom.
\label{fig:line}}
\end{center}
\end{figure}

\begin{figure}
\begin{center}
\includegraphics[scale=0.85]{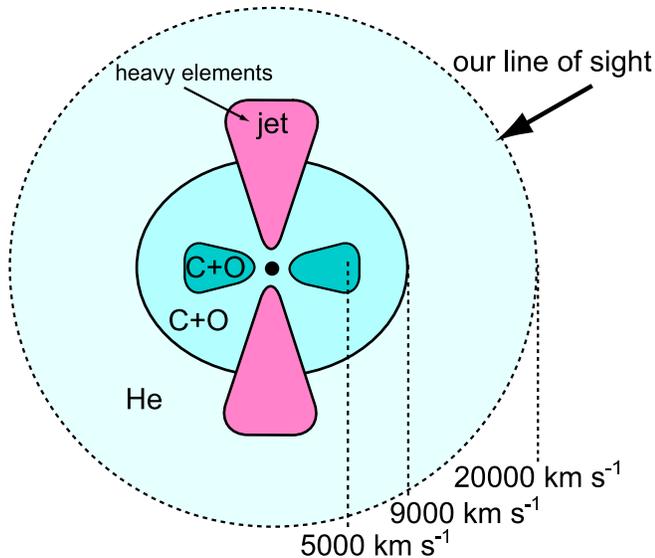}
\caption{
Schematic illustration of a bipolar explosion model for SN 2008D.
O-rich material forms a torus-like distribution.
The emssion line from the torus-like distribution has a 
double-peaked profile if it is seen from near the equatorial 
direction ($\gsim 50^{\circ}$ from the pole).
\label{fig:model}}
\end{center}
\end{figure}

A simple explanation for the double-peaked [\ion{O}{i}] 
profile is that 
oxygen has a torus-like distribution and 
our line of sight is near the plane of the torus
(\citealt{maeda02,mazzali05,maeda08,modjaz08}, but see also 
discussion by \citealt{milisavljevic09}).
The torus-like distribution of the presupernova elements,
such as oxygen and carbon,
can be realized in a bipolar explosion
(see Figure \ref{fig:model} for a schematic picture),
where more energy is deposited in the polar region, 
and nucleosynthesis takes place there
\citep[\eg][]{nagataki97,khokhlov99,maeda03nuc}.
A bipolar explosion has also been suggested for SN 2008D 
by polarimetric observations \citep{gorosabel08}.

Nebular emission profiles from bipolar models have been examined 
by \citet{maeda06nebular}, and compared  
to nebular spectra of various SNe Ib/c \citep{maeda08}.
In the right-hand panel of Figure \ref{fig:line},
the profiles calculated with a bipolar explosion model BP8
(highly aspherical model) by \citet{maeda06nebular} are shown for comparison.
The kinetic energy of the model is $2 \times 10^{51}$ erg
($f_v = 0.7$ defined in \citealt{maeda06nebular}), 
which is smaller than that estimated for SN 2008D 
\citep[\eg][]{tanaka0908D}.
This is because the models of \citet{maeda06nebular} 
include only the C+O core.
In SN 2008D, and in the models of \citet{tanaka0908D}, 
the He layer is present, and it contains the rest of the kinetic energy.

The line profile of SN 2008D is explained
most nicely by a viewing angle of $\sim 60-70^{\circ}$.
The double-peaked profile is seen
only when the viewing angle is $> 50^{\circ}$.
If the asphericity is smaller, this dividing angle is larger.
Since the viewing angle and degree of asphericity have a degenerate effect, 
we cannot argue that a large asphericity as in model BP8 
is preferred for SN 2008D.
However, it is true irrespective of the models that our 
line of sight should be 
$> 50^{\circ}$ from the polar direction.
If the angle is smaller, 
the [\ion{O}{i}] line would show a single-peaked
profile (see SNe 1998bw, 2004gq and 2002ap and models  
in Figure \ref{fig:line}).

\section{Conclusions}
\label{sec:con}

We have presented spectroscopic and photometric observations 
of SN 2008D at nebular phases.
Flux measurements of the narrow emission lines from the 
\ion{H}{ii} region at the site of the SN and the estimates of the 
local metallicity were also given.
In photometric observations, the SN was detected in $R$ and $I$.
The brightness at nebular phases is consistent
with the prediction of explosion models with an ejected 
\Nifs\ mass of 0.07 $\Msun$, which explain the light curve at early phases.

In the nebular spectrum taken at a sufficiently late phase,
the [\ion{O}{i}] line clearly shows a double-peaked profile,
while such a profile is not seen in the [\ion{Ca}{ii}] line.
The evolution of the profile from $t_{\rm exp} = 109$ days 
\citep{modjaz0808D} turned out not to be significant.
The double-peaked profile cannot be explained 
by a spherically symmetric explosion, and 
it strongly indicates that SN 2008D is an aspherical explosion.
The profile can be explained by 
a torus-like distribution of excited \ion{O}{i} viewed from the side
(Figure \ref{fig:model}).
Our line of sight is $> 50^{\circ}$ from the polar direction,
irrespective of the degree of asphericity.

\acknowledgments
We are grateful to the staff of the Subaru Telescope for their
kind support.
M.T. would like to thank Yusei Koyama for helpful discussion,
and the anonymous referee for useful comments.
M.T. is supported by the 
Japan Society for the Promotion of Science
Research Fellowship for Young Scientists.
This research has been supported in part by World Premier
International Research Center Initiative, MEXT,
Japan, and by the Grant-in-Aid for Scientific Research of the JSPS
(18104003, 18540231, 20540226, 20840007) and MEXT (19047004, 20040004).


\begin{thebibliography}{40}
\expandafter\ifx\csname natexlab\endcsname\relax\def\natexlab#1{#1}\fi

\bibitem[{{Appenzeller} {et~al.}(1998){Appenzeller}, {Fricke}, {F{\"u}rtig},
  {G{\"a}ssler}, {H{\"a}fner}, {Harke}, {Hess}, {Hummel}, {J{\"u}rgens},
  {Kudritzki}, {Mantel}, {Meisl}, {Muschielok}, {Nicklas}, {Rupprecht},
  {Seifert}, {Stahl}, {Szeifert}, \& {Tarantik}}]{appenzeller98}
{Appenzeller}, I., {et~al.} 1998, The Messenger, 94, 1

\bibitem[{{Asplund} {et~al.}(2004){Asplund}, {Grevesse}, {Sauval}, {Allende
  Prieto}, \& {Kiselman}}]{asplund04}
{Asplund}, M., {Grevesse}, N., {Sauval}, A.~J., {Allende Prieto}, C., \&
  {Kiselman}, D. 2004, \aap, 417, 751

\bibitem[{{Berger} \& {Soderberg}(2008)}]{berger08}
{Berger}, E., \& {Soderberg}, A.~M. 2008, GRB Coordinates Network, 7159, 1

\bibitem[{{Cardelli} {et~al.}(1989){Cardelli}, {Clayton}, \&
  {Mathis}}]{cardelli89}
{Cardelli}, J.~A., {Clayton}, G.~C., \& {Mathis}, J.~S. 1989, \apj, 345, 245

\bibitem[{{Chevalier} \& {Fransson}(2008)}]{chevalier08}
{Chevalier}, R.~A., \& {Fransson}, C. 2008, \apjl, 683, L135

\bibitem[{{Deng} \& {Zhu}(2008)}]{deng08}
{Deng}, J., \& {Zhu}, Y. 2008, GRB Coordinates Network, 7160, 1

\bibitem[{{Fransson} \& {Chevalier}(1989)}]{fransson89}
{Fransson}, C., \& {Chevalier}, R.~A. 1989, \apj, 343, 323

\bibitem[{{Gorosabel} {et~al.}(2008){Gorosabel}, {de Ugarte Postigo},
  {Castro-Tirado}, {Agudo}, {Jelinek}, {Leon}, {Augusteijn}, {Fynbo}, {Hjorth},
  {Michalowski}, {Xu}, {Ferrero}, {Kann}, {Klose}, {Rossi}, {Madrid},
  {LLorente}, {Bremer}, \& {Winters}}]{gorosabel08}
{Gorosabel}, J., {et~al.} 2008, submitted to ApJL (arXiv:0810.4333)

\bibitem[{{Kashikawa} {et~al.}(2002){Kashikawa}, {Aoki}, {Asai}, {Ebizuka},
  {Inata}, {Iye}, {Kawabata}, {Kosugi}, {Ohyama}, {Okita}, {Ozawa}, {Saito},
  {Sasaki}, {Sekiguchi}, {Shimizu}, {Taguchi}, {Takata}, {Yadoumaru}, \&
  {Yoshida}}]{kashikawa02}
{Kashikawa}, N., {et~al.} 2002, \pasj, 54, 819

\bibitem[{{Katz} {et~al.}(2009){Katz}, {Budnik}, \& {Waxman}}]{katz09}
{Katz}, B., {Budnik}, R., \& {Waxman}, E. 2009 (arXiv:0902.4708)

\bibitem[{{Khokhlov} {et~al.}(1999){Khokhlov}, {H{\"o}flich}, {Oran},
  {Wheeler}, {Wang}, \& {Chtchelkanova}}]{khokhlov99}
{Khokhlov}, A.~M., {H{\"o}flich}, P.~A., {Oran}, E.~S., {Wheeler}, J.~C.,
  {Wang}, L., \& {Chtchelkanova}, A.~Y. 1999, \apjl, 524, L107

\bibitem[{{Leibundgut} {et~al.}(1991){Leibundgut}, {Kirshner}, {Pinto},
  {Rupen}, {Smith}, {Gunn}, \& {Schneider}}]{leibundgut91}
{Leibundgut}, B., {Kirshner}, R.~P., {Pinto}, P.~A., {Rupen}, M.~P., {Smith},
  R.~C., {Gunn}, J.~E., \& {Schneider}, D.~P. 1991, \apj, 372, 531

\bibitem[{{Li}(2008)}]{lixl08}
{Li}, L.-X. 2008, \mnras, 388, 603

\bibitem[{{Li} \& {Filippenko}(2008)}]{liw08}
{Li}, W., \& {Filippenko}, A.~V. 2008, Central Bureau Electronic Telegrams,
  1202, 3

\bibitem[{{Maeda} {et~al.}(2008){Maeda}, {Kawabata}, {Mazzali}, {Tanaka},
  {Valenti}, {Nomoto}, {Hattori}, {Deng}, {Pian}, {Taubenberger}, {Iye},
  {Matheson}, {Filippenko}, {Aoki}, {Kosugi}, {Ohyama}, {Sasaki}, \&
  {Takata}}]{maeda08}
{Maeda}, K., {et~al.} 2008, Science, 319, 1220

\bibitem[{{Maeda} {et~al.}(2002){Maeda}, {Nakamura}, {Nomoto}, {Mazzali},
  {Patat}, \& {Hachisu}}]{maeda02}
{Maeda}, K., {Nakamura}, T., {Nomoto}, K., {Mazzali}, P.~A., {Patat}, F., \&
  {Hachisu}, I. 2002, \apj, 565, 405

\bibitem[{{Maeda} \& {Nomoto}(2003)}]{maeda03nuc}
{Maeda}, K., \& {Nomoto}, K. 2003, \apj, 598, 1163

\bibitem[{{Maeda} {et~al.}(2006){Maeda}, {Nomoto}, {Mazzali}, \&
  {Deng}}]{maeda06nebular}
{Maeda}, K., {Nomoto}, K., {Mazzali}, P.~A., \& {Deng}, J. 2006, \apj, 640, 854

\bibitem[{{Malesani} {et~al.}(2009){Malesani}, {Fynbo}, {Hjorth}, {Leloudas},
  {Sollerman}, {Stritzinger}, {Vreeswijk}, {Watson}, {Gorosabel},
  {Micha{\l}owski}, {Th{\"o}ne}, {Augusteijn}, {Bersier}, {Jakobsson},
  {Jaunsen}, {Ledoux}, {Levan}, {Milvang-Jensen}, {Rol}, {Tanvir}, {Wiersema},
  {Xu}, {Albert}, {Bayliss}, {Gall}, {Grove}, {Koester}, {Leitet}, {Pursimo},
  \& {Skillen}}]{malesani09}
{Malesani}, D., {et~al.} 2009, \apjl, 692, L84

\bibitem[{{Mazzali} {et~al.}(2007){Mazzali}, {Kawabata}, {Maeda}, {Foley},
  {Nomoto}, {Deng}, {Suzuki}, {Iye}, {Kashikawa}, {Ohyama}, {Filippenko},
  {Qiu}, \& {Wei}}]{mazzali07}
{Mazzali}, P.~A., {et~al.} 2007, \apj, 670, 592

\bibitem[{{Mazzali} {et~al.}(2005){Mazzali}, {Kawabata}, {Maeda}, {Nomoto},
  {Filippenko}, {Ramirez-Ruiz}, {Benetti}, {Pian}, {Deng}, {Tominaga},
  {Ohyama}, {Iye}, {Foley}, {Matheson}, {Wang}, \& {Gal-Yam}}]{mazzali05}
---. 2005, Science, 308, 1284

\bibitem[{{Mazzali} {et~al.}(2008){Mazzali}, {Valenti}, {Della Valle},
  {Chincarini}, {Sauer}, {Benetti}, {Pian}, {Piran}, {D'Elia}, {Elias-Rosa},
  {Margutti}, {Pasotti}, {Antonelli}, {Bufano}, {Campana}, {Cappellaro},
  {Covino}, {D'Avanzo}, {Fiore}, {Fugazza}, {Gilmozzi}, {Hunter}, {Maguire},
  {Maiorano}, {Marziani}, {Masetti}, {Mirabel}, {Navasardyan}, {Nomoto},
  {Palazzi}, {Pastorello}, {Panagia}, {Pellizza}, {Sari}, {Smartt},
  {Tagliaferri}, {Tanaka}, {Taubenberger}, {Tominaga}, {Trundle}, \&
  {Turatto}}]{mazzali08}
---. 2008, Science, 321, 1185

\bibitem[{{Milisavljevic} {et~al.}(2009){Milisavljevic}, {Fesen}, {Gerardy},
  {Kirshner}, \& {Challis}}]{milisavljevic09}
{Milisavljevic}, D., {Fesen}, R., {Gerardy}, C., {Kirshner}, R., \& {Challis},
  P. 2009 (arXiv:0904.4256)

\bibitem[{{Modjaz} {et~al.}(2008{\natexlab{a}}){Modjaz}, {Kewley}, {Kirshner},
  {Stanek}, {Challis}, {Garnavich}, {Greene}, {Kelly}, \&
  {Prieto}}]{modjaz08host}
{Modjaz}, M., {et~al.} 2008{\natexlab{a}}, \aj, 135, 1136

\bibitem[{{Modjaz} {et~al.}(2008{\natexlab{b}}){Modjaz}, {Kirshner}, {Blondin},
  {Challis}, \& {Matheson}}]{modjaz08}
{Modjaz}, M., {Kirshner}, R.~P., {Blondin}, S., {Challis}, P., \& {Matheson},
  T. 2008{\natexlab{b}}, \apjl, 687, L9

\bibitem[{{Modjaz} {et~al.}(2008{\natexlab{c}}){Modjaz}, {Li}, {Butler},
  {Chornock}, {Perley}, {Blondin}, {Bloom}, {Filippenko}, {Kirshner},
  {Kocevski}, {Poznanski}, {Hicken}, {Foley}, {Stringfellow}, {Berlind},
  {Barrado y Navascues}, {Blake}, {Bouy}, {Brown}, {Challis}, {Chen}, {de
  Vries}, {Dufour}, {Falco}, {Friedman}, {Ganeshalingam}, {Garnavich},
  {Holden}, {Illingworth}, {Liebert}, {Marion}, {Lee}, {Olivier}, {Olszewski},
  {Prochaska}, {Silverman}, {Smith}, {Starr}, {Steele}, {Stockton}, {Williams},
  \& {Wood-Vasey}}]{modjaz0808D}
{Modjaz}, M., {et~al.} 2008{\natexlab{c}}, submitted to ApJ (arXiv:0805.2201)


\bibitem[{{Nagataki} {et~al.}(1997){Nagataki}, {Hashimoto}, {Sato}, \&
  {Yamada}}]{nagataki97}
{Nagataki}, S., {Hashimoto}, M.-A., {Sato}, K., \& {Yamada}, S. 1997, \apj,
  486, 1026

\bibitem[{{Oke}(1990)}]{oke90}
{Oke}, J.~B. 1990, \aj, 99, 1621

\bibitem[{{Patat} {et~al.}(2001){Patat}, {Cappellaro}, {Danziger}, {Mazzali},
  {Sollerman}, {Augusteijn}, {Brewer}, {Doublier}, {Gonzalez}, {Hainaut},
  {Lidman}, {Leibundgut}, {Nomoto}, {Nakamura}, {Spyromilio}, {Rizzi},
  {Turatto}, {Walsh}, {Galama}, {van Paradijs}, {Kouveliotou}, {Vreeswijk},
  {Frontera}, {Masetti}, {Palazzi}, \& {Pian}}]{patat01}
{Patat}, F., {et~al.} 2001, \apj, 555, 900

\bibitem[{{Pettini} \& {Pagel}(2004)}]{pettini04}
{Pettini}, M., \& {Pagel}, B.~E.~J. 2004, \mnras, 348, L59

\bibitem[{{Soderberg} {et~al.}(2008){Soderberg}, {Berger}, {Page}, {Schady},
  {Parrent}, {Pooley}, {Wang}, {Ofek}, {Cucchiara}, {Rau}, {Waxman}, {Simon},
  {Bock}, {Milne}, {Page}, {Barentine}, {Barthelmy}, {Beardmore}, {Bietenholz},
  {Brown}, {Burrows}, {Burrows}, {Byrngelson}, {Cenko}, {Chandra}, {Cummings},
  {Fox}, {Gal-Yam}, {Gehrels}, {Immler}, {Kasliwal}, {Kong}, {Krimm},
  {Kulkarni}, {Maccarone}, {M{\'e}sz{\'a}ros}, {Nakar}, {O'Brien}, {Overzier},
  {de Pasquale}, {Racusin}, {Rea}, \& {York}}]{soderberg08}
{Soderberg}, A.~M., {et~al.} 2008, \nat, 453, 469

\bibitem[{{Sollerman} {et~al.}(1998){Sollerman}, {Leibundgut}, \&
  {Spyromilio}}]{sollerman98}
{Sollerman}, J., {Leibundgut}, B., \& {Spyromilio}, J. 1998, \aap, 337, 207

\bibitem[{{Tanaka} {et~al.}(2009){Tanaka}, {Tominaga}, {Nomoto}, {Valenti},
  {Sahu}, {Minezaki}, {Yoshii}, {Yoshida}, {Anupama}, {Benetti}, {Chincarini},
  {Valle}, {Mazzali}, \& {Pian}}]{tanaka0908D}
{Tanaka}, M., {et~al.} 2009, \apj, 692, 1131

\bibitem[{{Taubenberger} {et~al.}(2009){Taubenberger}, {Valenti}, {Benetti},
  {Cappellaro}, {Della Valle}, {Elias-Rosa}, {Hachinger}, {Hillebrandt},
  {Maeda}, {Mazzali}, {Pastorello}, {Patat}, {Sim}, \&
  {Turatto}}]{taubenberger09}
{Taubenberger}, S., {et~al.} 2009, MNRAS, in press (arXiv:0904.4632)

\bibitem[{{Th{\"o}ne} {et~al.}(2009){Th{\"o}ne}, {Micha{\l}owski}, {Leloudas},
  {Cox}, {Fynbo}, {Sollerman}, {Hjorth}, \& {Vreeswijk}}]{thoene08}
{Th{\"o}ne}, C.~C., {Micha{\l}owski}, M.~J., {Leloudas}, G., {Cox}, N.~L.~J.,
  {Fynbo}, J.~P.~U., {Sollerman}, J., {Hjorth}, J., \& {Vreeswijk}, P.~M. 2009,
  \apj, 698, 1307

\bibitem[{{Tomita} {et~al.}(2006){Tomita}, {Deng}, {Maeda}, {Yoshii}, {Nomoto},
  {Mazzali}, {Suzuki}, {Kobayashi}, {Minezaki}, {Aoki}, {Enya}, \&
  {Suganuma}}]{tomita06}
{Tomita}, H., {et~al.} 2006, \apj, 644, 400

\bibitem[{{Valenti} {et~al.}(2008{\natexlab{a}}){Valenti}, {Benetti},
  {Cappellaro}, {Patat}, {Mazzali}, {Turatto}, {Hurley}, {Maeda}, {Gal-Yam},
  {Foley}, {Filippenko}, {Pastorello}, {Challis}, {Frontera}, {Harutyunyan},
  {Iye}, {Kawabata}, {Kirshner}, {Li}, {Lipkin}, {Matheson}, {Nomoto}, {Ofek},
  {Ohyama}, {Pian}, {Poznanski}, {Salvo}, {Sauer}, {Schmidt}, {Soderberg}, \&
  {Zampieri}}]{valenti0803jd}
{Valenti}, S., {et~al.} 2008{\natexlab{a}}, \mnras, 383, 1485

\bibitem[{{Valenti} {et~al.}(2008{\natexlab{b}}){Valenti}, {Turatto},
  {Navasardyan}, {Benetti}, \& {Cappellaro}}]{valenti0808D}
{Valenti}, S., {Turatto}, M., {Navasardyan}, H., {Benetti}, S., \&
  {Cappellaro}, E. 2008{\natexlab{b}}, GRB Coordinates Network, 7163, 1

\bibitem[{{Xu} {et~al.}(2008){Xu}, {Watson}, {Fynbo}, {Fan}, {Zou}, \&
  {Hjorth}}]{xu08}
{Xu}, D., {Watson}, D., {Fynbo}, J., {Fan}, Y., {Zou}, Y.-C., \& {Hjorth}, J.
  2008, in COSPAR, Plenary Meeting, Vol.~37, 37th COSPAR Scientific Assembly,
  3512--+

\bibitem[{{Yoshii} {et~al.}(2003){Yoshii}, {Tomita}, {Kobayashi}, {Deng},
  {Maeda}, {Nomoto}, {Mazzali}, {Umeda}, {Aoki}, {Doi}, {Enya}, {Minezaki},
  {Suganuma}, \& {Peterson}}]{yoshii03}
{Yoshii}, Y., {et~al.} 2003, \apj, 592, 467

\end{thebibliography}
\end{document}